\documentclass[preprint,11pt,aps]{revtex4-1}
\usepackage[utf8]{inputenc}
\usepackage[T1]{fontenc}
\usepackage{bm}
\usepackage{amsmath,ulem}
\usepackage{amsfonts}
\usepackage{amssymb}
\usepackage{graphicx}
\usepackage{siunitx}
\usepackage{xcolor}

\definecolor{myBlue}{rgb}{0.1,0.1,0.7}

\newcommand{\Runit}{\si{\micro \metre}}

\newcommand{\Vunit}{\si{\micro\metre \,\second^{-1}}}

\newcommand{\Punit}{\si{\milli\watt}}

\newcommand{\Qunit}{\si{\micro\watt \, (\micro \meter)^{-1}}}

\begin{document}

\title{Programmable Hydrodynamics of Active Particles}

\author{Lisa Rohde}
    \affiliation{Molecular Nanophotonics Group, Peter Debye Institute for Soft Matter Physics, Leipzig University, 04103 Leipzig, Germany}
\author{Gordei Anchutkin}
    \affiliation{Molecular Nanophotonics Group, Peter Debye Institute for Soft Matter Physics, Leipzig University, 04103 Leipzig, Germany}
\author{Viktor Holubec}
    \affiliation{Department of Macromolecular Physics, Faculty of Mathematics and Physics,Charles University, 18000 Prague, Czech Republic}
\author{Frank Cichos}
    \email{cichos@physik.uni-leipzig.de}
    \affiliation{Molecular Nanophotonics Group, Peter Debye Institute for Soft Matter Physics, Leipzig University, 04103 Leipzig, Germany}

\begin{abstract}
Self-propelled microparticles create flow fields that determine how they interact with surfaces, external flows, and each other. These flow fields fall into distinct classes—pushers, pullers, and neutral swimmers—each exhibiting fundamentally different collective behaviors. In all existing synthetic systems, this hydrodynamic character is permanently set during fabrication, making it impossible to explore how adaptive switching between these classes might enable new functions or emergent phenomena.

Here we demonstrate that the hydrodynamic character of a microswimmer can be programmed and switched on demand. Using patterned laser heating of surface-bound nanoparticles, we create tailored temperature gradients that drive controllable boundary flows at the particle surface. By changing the illumination pattern in real time, we dynamically transform the swimmers flow field continuously tuning from pusher to puller, while the particle continues to swim. Flow measurements confirm quantitative agreement with theory and allow us to simultaneously track how symmetry, power consumption, and efficiency change across modes.
This control over hydrodynamic modes opens experimental access to questions that have remained largely theoretical: How do adaptive swimmers respond to crowding or confinement? Can mixtures with tunable pusher-puller ratios reveal new collective states? Our approach provides a platform to address these questions and explore the morphological developments of active matter systems under external physical constraints.
\end{abstract}

\maketitle

\section*{Introduction}\label{intro}
Micron-sized self-propelled particles—from swimming bacteria and motile algae to synthetic colloidal swimmers—generate long range hydrodynamic flow fields that extend far beyond the particle itself \cite{lighthill1952squirming,blake1971spherical,bechinger2016active}. At these length scales, where Reynolds numbers are small ($\mathrm{Re} \ll 1$) and viscous forces dominate, swimmers are classified by their far-field flow structure as pushers (bacteria like \textit{E. coli} that push fluid from the rear \cite{drescher2011fluid,hu2015modelling, lopez2014dynamics}), pullers (algae like \textit{Chlamydomonas} that pull fluid from the front \cite{klindt2015flagellar}), or neutral swimmers \cite{qi2025unravel}. A fundamental prediction of hydrodynamic theory is that spherical neutral swimmers achieve optimal swimming efficiency by eliminating slowly-decaying force dipole contributions to viscous dissipation \cite{daddi2021optimal,michelin2010efficiency,nasouri2021minimum, stone1996propulsion}. This result—that the neutral mode minimizes energy dissipation—has become a cornerstone of active matter theory and is frequently invoked to explain performance optimization in biological and synthetic swimmers. However, the theoretical derivation assumes arbitrary control over surface boundary conditions, unconstrained by physical limitations of the propulsion mechanism.

Real swimming systems—whether biological or synthetic—operate under physical constraints that limit achievable surface flows. Bacteria cannot arbitrarily program their flagellar arrangements \cite{constantino2018bipolar, ishikawa2009suspension}, chemically-powered Janus particles face asymmetric catalyst placement constraints \cite{zhang2023janus,ibrahim2017multiple,popescu2016self, theurkauff2012dynamic}, and thermophoretic swimmers require specific heating distributions \cite{bickel2014flow,jiang2010active,fraenzl2021fully}. Despite the ubiquity of such constraints and theoretical suggestions that they fundamentally alter performance optima,  no experimental system has been able to test this prediction. This challenge is formidable. Validating how constraints affect optimal swimming modes requires the ability to systematically vary the hydrodynamic character while quantifying both flow fields and energetic costs—capabilities absent in existing synthetic systems where the swimming mode is permanently fixed during fabrication \cite{shields2017evolution,yan2016reconfiguring,ketzetzi2025active}.

While recent advances have achieved controlled switching of propulsion direction and activity states \cite{vutukuri2020light,lee2019directed,ketzetzi2025active}, dynamic reconfiguration of the flow field signature itself—transforming between pusher, puller, and neutral modes—has remained elusive. The inability to reconfigure hydrodynamic modes has broader consequences for active matter research. Understanding how physical constraints shape morphological adaptation is essential for interpreting biological swimming strategies \cite{klindt2015flagellar,drescher2009dancing, zayed2025swimmer,singh2020mechanical} and for developing design principles for synthetic active matter \cite{grober2023unconventional, chakraborty2022self,grauer2021active,ben2023morphological}. If constraints shift performance optima away from hydrodynamic predictions, then evolutionary pressures and engineering optimization must account for the coupling between propulsion mechanism and hydrodynamic signature. Testing this requires precise, real-time control over surface boundary conditions—a capability that would enable experimental exploration of constrained optimization landscapes.

Here, we demonstrate real-time reconfiguration of hydrodynamic flow fields in synthetic microswimmers, enabling dynamic switching between pusher, puller, and neutral modes through programmable surface boundary conditions. Using spatially structured laser heating on gold nanoparticle-decorated microparticles, we achieve quantitative control over surface temperature distributions that determine the hydrodynamic signature—as validated by direct flow field measurements across the complete swimming mode spectrum. This platform enables the first experimental determination of swimming efficiency as a function of hydrodynamic mode under realistic physical constraints. We find a striking result: pusher swimmers, not neutral swimmers as pure hydrodynamic theory predicts, achieve maximum efficiency under the constraint that surface heating must remain positive everywhere. This demonstrates that physical constraints on control parameters fundamentally alter performance optima, shifting the optimal mode from neutral to pusher. Importantly, this principle extends beyond thermophoretic swimmers to any system where the propulsion mechanism imposes constraints on achievable surface flows. Our results establish a framework for understanding morphological optimization in active matter, where performance landscapes are determined by the interplay between hydrodynamic efficiency and mechanism-specific constraints—a perspective that opens new directions for exploring morphological intelligence and adaptive strategies in both biological and synthetic active systems.

\begin{figure}[h!]
    \centering
    \includegraphics[width=1\linewidth]{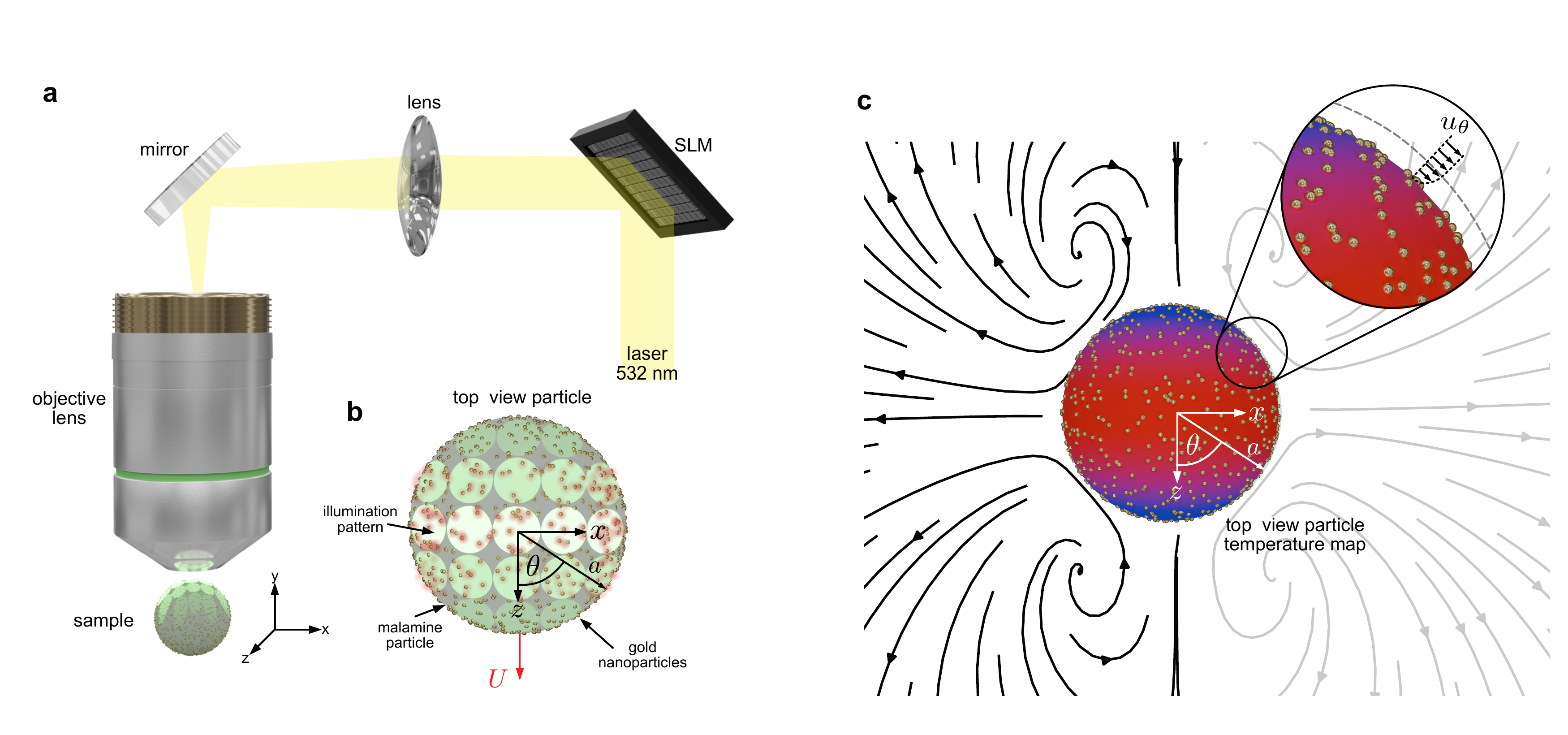}
    \caption{\textbf{Experimental realization of hydrodynamic reconfiguration} \textbf{a} Reconfiguring hydrodynamic flows of a microswimmer is achieved by selectively illuminating different regions on the particle's surface via a spatial light modulator (SLM). \textbf{b} The microswimmer is a melamine sphere with radius $a = 2.2\,$µm that is uniformly covered with gold nanoparticles that behave as heat sources when illuminated with laser light at $532\,$nm. The presented structured light pattern induces a heating at the equator of the particle at $\theta = \pi$ which will lead to self-propulsion $U$ along $z$. The angle $\theta$ is defined as the polar angle to the positive z-axis across the surface of the microswimmer. \textbf{c} All heated nanoparticles create a tangential temperature gradient which generates a thermo-osmotic boundary flow $u_\theta$ indicated by the black arrows. This boundary flow creates a hydrodynamic flow field which is characteristic for the selected swimming mode and resembles here a force dipole field.
    }
    \label{fig:Figure1}
\end{figure}

\subsection*{Precise control of surface temperature patterns}
Reconfiguring the hydrodynamic flow morphology of a microswimmer requires precise manipulation of the hydrodynamic boundary conditions on the particle's surface. We achieve this by using a special type of thermophoretic swimmers consisting of melamine microspheres decorated with plasmonic gold nanoparticles whose interfacial flow fields are programmed through structured laser illumination using a spatial light modulator (SLM, Fig.~\ref{fig:Figure1}a). When illuminated at \SI{532}{\nano\meter} near their plasmon resonance, the nanoparticles act as local point-like heat sources \cite{Jiang2010}. The sparse coverage (approximately 10\% surface coverage) maintains a low thermal conductivity along the surface, enabling creation of well-defined temperature patterns by selectively illuminating different surface regions.
Fig.~\ref{fig:Figure1}b displays this principle with a particular illumination pattern on the particle's surface that generates a heated region at its equator on the illumination side of the particle. For simplicity, we neglect refraction effects and assume a corresponding illumination pattern on the opposite side of the particle. The resulting temperature distribution is indicated in Fig.~\ref{fig:Figure1}c, with temperature gradients leading to thermo-osmotic boundary flows $u_\theta$ as indicated by the black arrows. These boundary flows generate a bulk hydrodynamic flow field sketched by the black streamlines in Fig.~\ref{fig:Figure1}c.

To induce a specific hydrodynamic flow morphology we can follow the squirmer framework \cite{lighthill1952squirming,blake1971spherical}, parameterizing the tangential slip velocity $u_\theta(\theta)$ at the particle surface ($r=a$) as
\begin{equation}\label{eq:squirmer_slip}
    u_\theta(\theta) = B_1\sin\theta + B_2\sin\theta\cos\theta\,,
\end{equation}
where $B_1$ drives propulsion and $B_2$ sets the far-field flow character. The dimensionless squirmer parameter $\beta = B_2/|B_1|$ classifies swimmers as pushers ($\beta <0$, extensile), pullers ($\beta > 0$, contractile), or neutral ($\beta = 0$, pure source dipole). Equation~\ref{eq:squirmer_slip} corresponds to the truncated Legendre polynomial expansion of the interfacial flow field $u_\theta(\theta)$ \cite{bickel2014flow} (see Supplementary Note 1).

These interfacial flows are generated by thermo-osmotic boundary flows that are themselves driven by interfacial temperature fields \cite{bregulla2016thermo}. The temperature fields arise from a surface heat source density $q(\theta)$ determined by the laser intensity pattern heating the gold nanoparticles. Exploiting the axial symmetry, we expand $q(\theta)$ in Legendre polynomials and truncate at second order to capture the essential physics:

\begin{equation}
q(\theta)= q_0 + q_1 \cos\theta +q_2 \frac{1}{2}(3\cos^2\theta-1)\,.\label{eq:q_modes}
\end{equation}

Since the heat source density scales linearly with incident intensity, the expansion coefficients directly determine the illumination pattern geometry. With $\cos\theta = z/a$ relating the polar angle to the axial coordinate $z$, these coefficients correspond to constant ($q_0$), linear ($q_1$), and quadratic ($q_2$) intensity profiles along the propulsion axis $z$. These heat source densities create temperature gradients that drive thermo-osmotic slip velocity $u_s(\theta) = (\chi/T_0)(1/a)\partial T(\theta)/\partial \theta$, where $\chi$ is the thermo-osmotic mobility characterizing the liquid-solid interfacial interaction \cite{Bregulla2019}. Solving the heat diffusion equation and matching to the desired squirmer modes (Eq.~\ref{eq:squirmer_slip}) establishes a direct correspondence: $q_1 \propto B_1$ and $q_2 \propto B_2$, enabling programming of any target squirmer parameter $\beta$ by adjusting the relative amplitudes $q_2/q_1$ (Supplementary Note 1 for detailed derivation).

\begin{figure}[h!]
    \centering
    \includegraphics[width=1\linewidth]{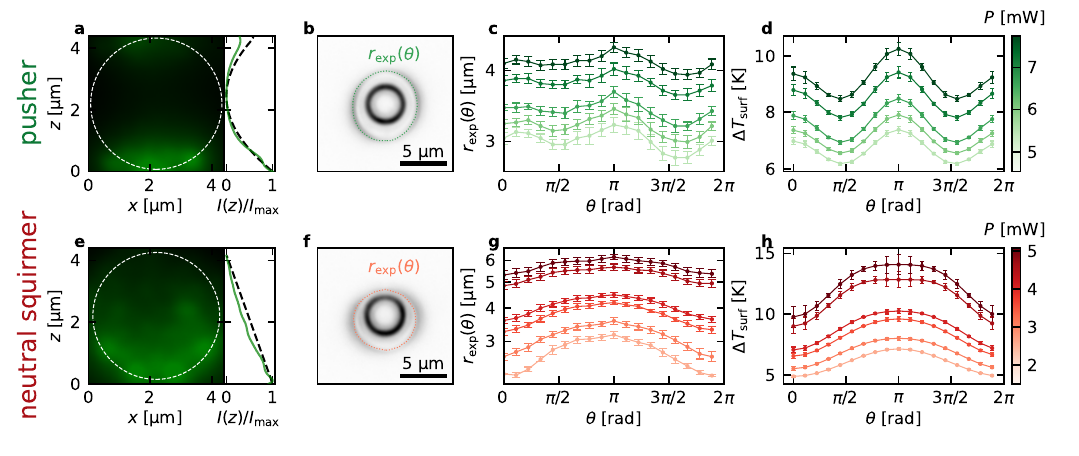}
    \caption{\textbf{Surface temperature control by precise illumination patterns} \textbf{a,e} We precisely shape the laser field using a spatial light modulator (SLM) to generate distinct illumination patterns, as shown here for a pusher with $\beta = -1$ and a neutral squirmer with $\beta = 0$. The measured intensity line profiles (green) agree well with the theoretical predictions from Eq.~\ref{eq:q_modes} (black), demonstrating accurate reproduction of the expected profiles. We obtain the microscopy images by exploiting the fluorescence of a dye at $\lambda = 532\,$nm. \textbf{b,f} The phase boundary of a liquid crystal forms when heating the microswimmer above the critical temperature of $308\,$K, revealing the asymmetric temperature profile on the surface of a swimmer, presented here for $\beta =-1$ and $\beta = 0$. The isotherms are highlighted by dashed lines defining the radius $r_\mathrm{exp}(\theta)$. \textbf{c,g} The angular dependence of $r_\mathrm{exp}(\theta)$, measured for different laser powers $P$ in the sample plane, confirm the precise control of the surface illumination. \textbf{d,h} The asymmetry of the isotherms ($r_\mathrm{exp}(\theta)$) gives information about the heat source density $q(\theta)$ on the particle's surface which can be used to reconstruct the surface temperature difference $\Delta T_\mathrm{surf}(\theta)$ as shown here for different heating powers $P$. Error bars represent the standard error of the mean calculated from multiple measurements. }
    \label{fig:Figure2}
\end{figure}

Figure~\ref{fig:Figure2}a,e shows this programmable control for two representative modes: a pusher ($\beta = -1$) and neutral squirmer ($\beta = 0$). The intensity patterns, measured using fluorescence of a dye excited at $\lambda = \SI{532}{\nano\meter}$, reveal the expected linear and quadratic profiles along the propulsion axis. Achieving such precise spatial control across the
\SI{4.4}{\micro\meter} diameter particle—only $\sim8$ wavelengths—demonstrates the remarkable
optical precision of our approach. In Fig.~\ref{fig:Figure2}a,e the intensity line profiles (green curves) match theoretical predictions (black curves) within 10\%, highlighting reproducible control over illumination patterns (Supplementary Fig.~S2 and S3, Supplementary Note 2).

We validate the resulting surface temperature distribution (Eq.~S6) by immobilizing particles in a liquid crystal (5CB) and exploiting the nematic-isotropic phase transition at the critical temperature (Supplementary Note 3). The phase boundary, visible under dark-field illumination due to refractive index mismatch between nematic and isotropic phase \cite{horn1978refractive}, reveals the asymmetric temperature profile as shown in Fig~\ref{fig:Figure2}b,f for the pusher and neutral squirmer. The isotherm at the phase transition temperature $T_\mathrm{PT} = 308\,$K (colored dashed line) defines a radius $r_\mathrm{exp}(\theta)$ that maps the asymmetric heating. Figure~\ref{fig:Figure2}c,g shows $r_\mathrm{exp}(\theta)$ for different laser powers, confirming high-precision control (see also Supplementary Fig.~S4). From $r_\mathrm{exp}(\theta)$, we directly extract the heat source coefficients $q_0$, $q_1$, and $q_2$ from Eq.~\ref{eq:q_modes}, enabling quantitative reconstruction of the surface temperature difference $\Delta T_\mathrm{surf}(\theta)$ (Supplementary Note 3 and Supplementary Fig.~S5, S6). Fig.~\ref{fig:Figure2}d,h shows $\Delta T_\mathrm{surf}(\theta)$ for various laser input powers-measured in the sample plane--for the two swimming modes. While the neutral squirmer reaches a maximum temperature increment of $14\,$K at an input power of $P=1.5\,\si\Punit$, the pusher's maximum surface temperature increment remains below $10\,$K even at substantially higher input powers, indicating differences in absorbed power between the swimming modes. The ability to precisely control surface temperature distributions is thus validated through both intensity measurements and direct liquid crystal thermography.

\subsection*{Experimental demonstration of hydrodynamic mode reconfiguration}
\begin{figure}[!h]
    \centering
    \includegraphics[width=0.9\linewidth]{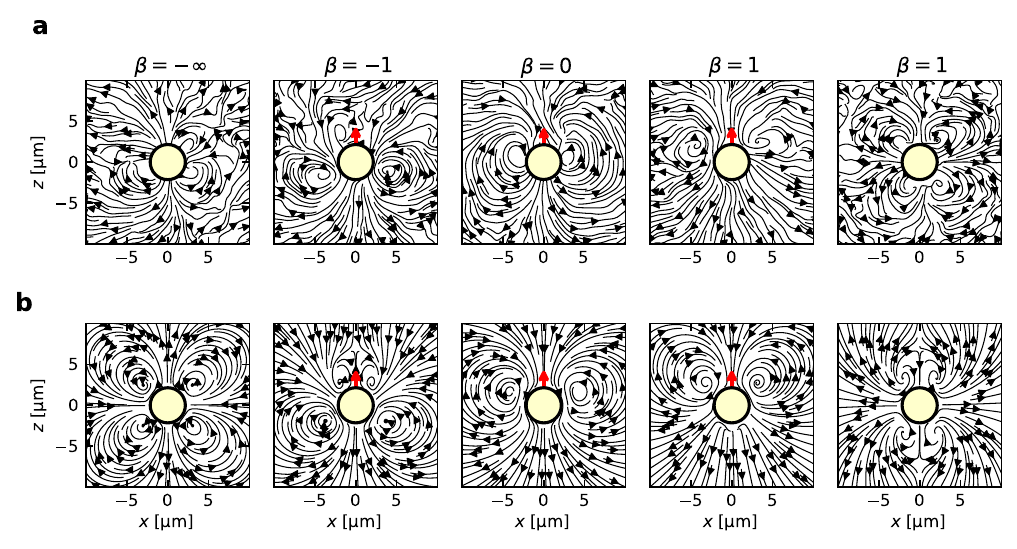}
    \includegraphics[width=0.82
    \linewidth]{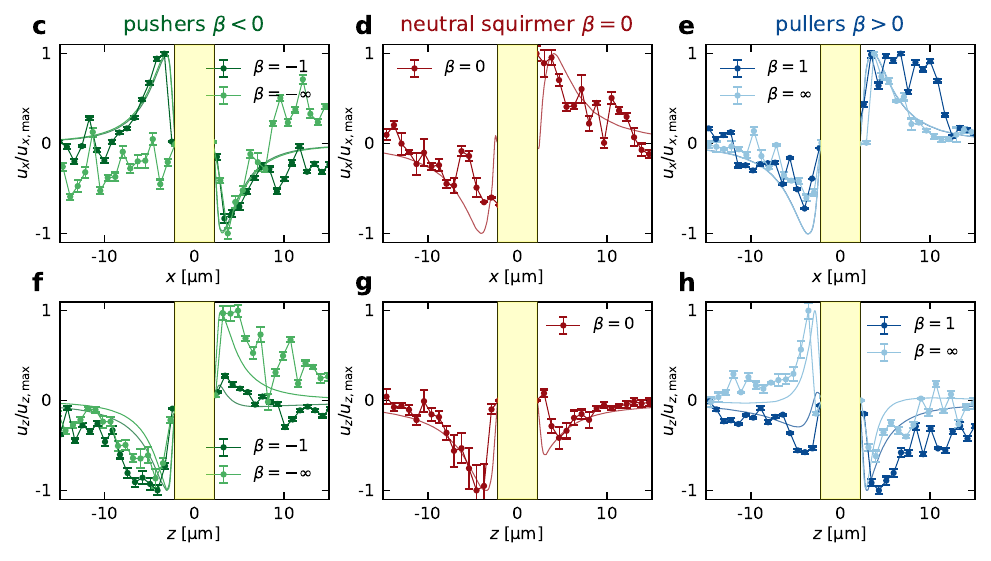}
    \caption{\textbf{Reconfigurable propulsion modes through hydrodynamic flow fields}  \textbf{a} Experimental flow fields for various swimming modes $\beta$, visualized using particle image velocimetry (PIV). The motion of tracer particles is represented as streamlines, with the red arrow indicating the propulsion direction. \textbf{b} Simulated flow fields are presented. The excellent agreement between experimental and simulated flow fields demonstrates the precise control over boundary flows achieved in the experiments, enabling reconfigurable propulsion modes through controlled variation of the illumination patterns. \textbf{c,d,e} While the normalized $x$-component of the velocity $u_x/u_{x,\mathrm{max}}$ is symmetric around $x$, the $z$-component $u_z/u_{z,\mathrm{max}}$ (\textbf{f,g,h}) indicates the polar asymmetry of the hydrodynamic flow fields. The measured velocity components (dotted lines) closely follow the decay of simulated velocity components (solid lines), demonstrating quantitative agreement between experiment and simulation. The swimmer's surface is marked with yellow color. Error bars represent the standard error of the mean calculated from multiple measurements.}
    \label{fig:Figure3}
\end{figure}

The programmable illumination patterns translate directly into reconfigurable hydrodynamic signatures. Figure \ref{fig:Figure3}a shows flow fields for immobilized microswimmers spanning the complete range of propulsion modes—from negative shaker ($\beta = -\infty$) through pusher ($\beta = -1$), neutral squirmer ($\beta = 0$), and puller ($\beta = 1$) to positive shaker ($\beta = \infty$) visualized using particle image velocimetry (PIV) with gold nanoparticle tracers. The neutral squirmer exhibits a pure source dipole flow field. At $\beta = \infty$, where the propulsive mode amplitude vanishes, only the force dipole signature remains.

Across all measured values of $\beta$, the experimental flow fields qualitatively match simulated results that account for confinement effects due to substrate boundaries and immobilization constraints which give rise to an additional Stokeslet (force-monopole) contribution to the flow field. (Fig.~\ref{fig:Figure3}b; Supplementary Note 4).  This confirms that varying only the illumination pattern reconfigures the swimmer's complete hydrodynamic signature without any mechanical modification.

The velocity field components (Fig.~\ref{fig:Figure3}c-h) reveal the underlying flow structure. The $x$-component remains symmetric about the $x$-axis with no variation in $|\beta|$ (Fig.~\ref{fig:Figure3}c-e), while the $z$-component shows more complex behavior (Fig.~\ref{fig:Figure3}f-h). For larger $|\beta|$ values, as in pushers, the $z$-component magnitude decreases, indicating reduced fluid expulsion from the swimmer's front and smaller flow vortices. Experimental velocity components (lines with markers and errorbars) closely follow simulated trends (solid lines) throughout with their characteristic distance dependence.

\subsection*{Moving swimmers and their efficiencies}
\begin{figure}[h!]
    \centering
    \includegraphics[width=0.65\linewidth]{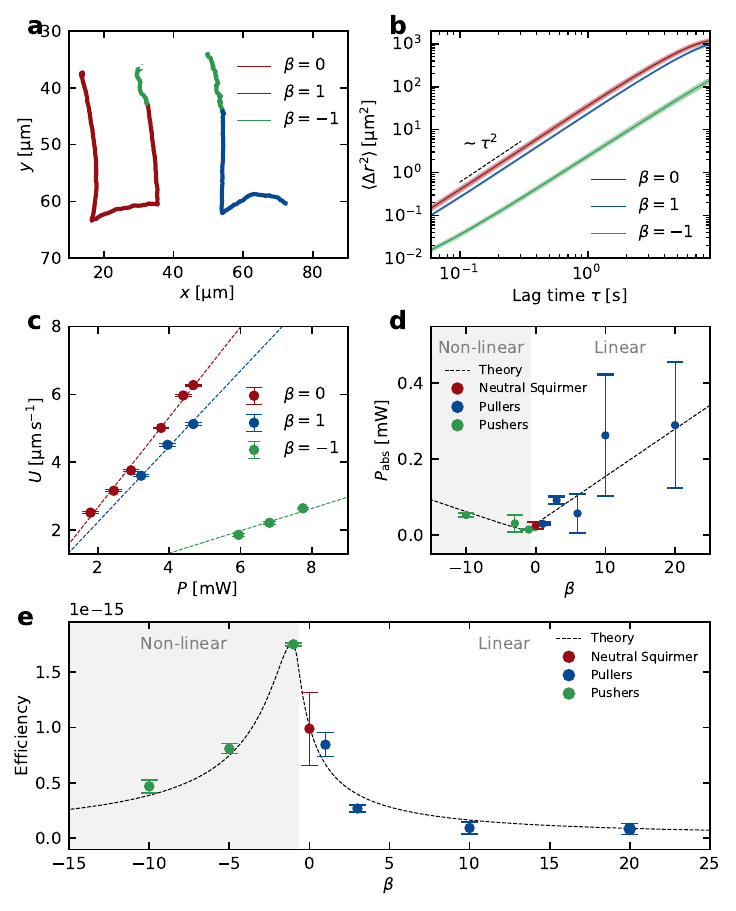}
    \caption{\textbf{Swimming efficiencies} \textbf{a} The trajectory of an active particle undergoing sequential transitions between distinct propulsion modes — from neutral squirmer (red) to pusher (green), and subsequently to puller (blue) — demonstrates the capability for instantaneous reconfiguration. \textbf{b} Mean squared displacement (MSD) as a function of lag time $\tau$ for a neutral squirmer ($\beta = 0$), puller ($\beta = 1$), and pusher ($\beta = -1$) at the same laser power of $P = 4.7\, \si{\Punit}$ scales with $\tau^2$, indicating ballistic motion of the active particle. \textbf{c} The velocity of the swimmers increases linearly with laser power $P$. The neutral squirmer reaches velocity values three times larger than the pusher. The dashed lines show fits of the form $U = mP$. \textbf{d} The absorbed power $P_\mathrm{abs}$ as a function of squirmer parameter $\beta$ for constant $q_1=- 0.4\,\si{\Qunit}$ indicates that the pusher with $\beta \approx -1$ exhibits the lowest absorbed power. \textbf{e} The experimentally determined efficiencies (symbols) for different swimming modes align well with the theoretical model (dashed line), demonstrating that the pusher with $\beta\approx -1$ is the most efficient swimmer and further validating the control and characterization of the propulsion modes. The non-linear and linear regimes are indicated by the gray and white backgrounds, respectively. Error bars represent the standard error of the mean calculated from multiple measurements.}
    \label{fig:Figure4}
\end{figure}

The reconfigurable flow fields translate into programmable swimming behavior for freely moving particles. Suspending swimmers in a thin liquid film and dynamically varying the illumination pattern enables real-time switching between propulsion modes. We therefore track the position of the swimmers within each frame and reposition and reorient the pattern correspondingly \cite{fraenzl2021fully}. Figure~\ref{fig:Figure4}a shows a single swimmer transitioning from neutral squirmer (red) to pusher (green) to puller (blue) during continuous motion, demonstrating instantaneous reconfigurability without mechanical modification (Supplementary Movies 1 and 2).

The three modes exhibit at first glance strikingly different swimming characteristics. All show quasi-ballistic motion with a mean squared displacement scaling as $\tau^2$ (Fig.~\ref{fig:Figure4}b), confirming active self-propulsion \cite{howse2007self}. The rotational diffusion of the particle is irrelevant for the dynamics as the symmetry of the pattern defines the propulsion direction not the geometry. However, the propulsion velocities differ dramatically. At a constant incident laser power ($P = 4.7\,\si{\Punit}$), the neutral squirmer reaches $6\,\si{\Vunit}$, three times faster than the pusher. Velocities scale linearly with laser power for all modes (Fig.~\ref{fig:Figure4}c), but the velocity-to-power ratios vary significantly, suggesting mode-dependent efficiencies as expected from literature \cite{daddi2021optimal, nasouri2021minimum}.

These strong velocity differences raise the question which mode swims most efficiently? While the neutral squirmer moves fastest and is expected to be the most efficient \cite{daddi2021optimal,nasouri2021minimum}, speed alone may not determine efficiency. Following Lighthill's model \cite{lighthill1952squirming}, we calculate swimming efficiency as the ratio of mechanical output power (Stokes drag: $F_\mathrm{stokes} U = 6\pi\eta a U^2$) to absorbed light power heating the particle $P_\mathrm{abs}$ \cite{bregulla2015size}:
\begin{align}
    \epsilon = \frac{P_\text{out}}{P_\text{in}} = \frac{6 \pi \eta a U^2}{P_\mathrm{abs}}\,.
    \label{eq:efficiency_main}
\end{align}
Unlike virtually all other phoretic swimmer experiments, our experimental system allows direct determination of the absorbed power $P_\mathrm{abs}$, i.e. the power that generates the temperature field. This power corresponds to the value of the constant heat source density $q_0$ term multiplied by particle surface area (Supplementary Note 5).

Extracting the heat source coefficients $q_0$, $q_1$, and $q_2$ from the measured temperature profiles (Fig.~\ref{fig:Figure2}c), we determine $P_\mathrm{abs}$ for each mode at constant propulsion velocity (fixed $q_1 = -0.4\,\si{\Qunit}$). Figure~\ref{fig:Figure4}d reveals that the absorbed power exhibits a pronounced minimum at $\beta \approx -1$ (pusher mode), not at the fastest-swimming neutral squirmer. This finding is independently confirmed by surface temperature measurements showing the pusher achieves the lowest $\Delta T_\mathrm{surf}$ (Supplementary Fig.~S5d).

The corresponding efficiencies (Fig.~\ref{fig:Figure4}e) display a sharp maximum at $\beta \approx -1$ in the experimentally determined values. They are as expected for many phoretic swimmers quite low and on the order of $10^{-15}$, indicating that a substantial amount of energy is used to maintain the gradients rather then propelling the particle. Remarkably though, the pusher despite swimming three times slower than the neutral squirmer emerges as the most efficient mode, requiring minimal thermal power input for a given propulsion velocity. This counterintuitive result, where the slowest swimmer proves most efficient, points to fundamental physical constraints governing thermophoretic propulsion that differ from purely hydrodynamic considerations.

\section*{Discussion}\label{discussion}
In this study, we demonstrate that real-time reconfiguration of hydrodynamic flow fields enables reversible switching between microparticles' swimming modes—from pullers to pushers and neutral squirmers—revealing that pusher swimmers achieve the highest swimming efficiency. Reconfigurability is achieved through precise energy manipulation of boundary flows of microswimmers. Our central finding—that pushers with $\beta \approx -1$ maximize thermophoretic swimming efficiency—stands in marked contrast to purely hydrodynamic predictions. Standard squirmer theory identifies the neutral squirmer ($\beta = 0$) as optimal, as it generates only a rapidly-decaying source dipole flow field, minimizing viscous dissipation \cite{daddi2021optimal, daddi2023minimum, michelin2010efficiency,nasouri2021minimum}. Similarly, size and shape-optimization studies focus on geometric considerations to tune efficiency while treating surface activity distributions as independently tunable parameters \cite{sabass2010efficiency,bregulla2015size,michelin2010efficiency}.

The resolution of this apparent paradox lies in recognizing that physical constraints fundamentally alter the optimization landscape. For thermophoretic swimmers, the requirement that heat source density remain positive everywhere on the particle surface, $q(\theta) \geq 0$ for all $\theta$—introduces a constraint that couples all multipole coefficients in the surface heat distribution. This coupling is absent in idealized models where surface slip velocities can be prescribed arbitrarily. While previous work has explored how geometric shape affects efficiency through modified resistance tensors \cite{daddi2023minimum}, our results reveal that realizability constraints on the driving fields themselves impose a more fundamental limitation: they determine not only the magnitude of efficiency but also which swimming mode is optimal.

The constraint $q(\theta) \geq 0$ forces a specific relationship between the monopole heating $q_0$, the propulsion-driving dipole $q_1$, and the mode-selecting quadrupole $q_2$. To maintain positive heating everywhere, the monopole term must compensate for the minimum of the combined dipole-quadrupole distribution, creating two distinct regimes (linear: $q_2 < -q_1/3$; nonlinear: $q_2 > -q_1/3$). The resulting swimming efficiencies in these regimes are (see Supplementary Note 5):
\begin{align}
    \epsilon_\mathrm{lin} & =\frac{2 \eta \chi^2 }{3a T_0^2(\kappa_\text{in} + 2\kappa_\text{out})^2}\frac{q_1^2}{(-q_1-q_2)}\,,\label{eq:efficiency_lin}\\
    \epsilon_\mathrm{non-lin} &=\frac{4\eta \chi^2}{3a T_0^2 (\kappa_\text{in} + 2\kappa_\text{out})^2}\frac{q_1^2}{\left(\frac{q_1^2}{3q_2} + q_2\right)}\,,
    \label{eq:efficiency_non_lin}
\end{align}
where $\eta$ is the fluid viscosity, $\chi$ is the thermo-osmotic mobility, $T_0$ is the ambient temperature, $a$ is the particle radius, and $\kappa_\text{in}$ and $\kappa_\text{out}$ are the thermal conductivities inside and outside the particle. These expressions reveal how the constraint-induced coupling manifests: efficiency depends not on $q_1$ and $q_2$ independently, but on their ratio, which determines $\beta = B_2/|B_1| \propto q_2/q_1$. Optimizing efficiency with respect to this ratio yields $\beta^* = -\sqrt{3} \frac{\kappa_\mathrm{in} + 2\kappa_\mathrm{out}}{2\kappa_\mathrm{in} + 3\kappa_\mathrm{out}} \approx -1$ for typical material parameters, precisely where we observe maximum experimental efficiency (Fig.~\ref{fig:Figure4}e).

Importantly, this $\beta \approx -1$ optimum is not an artifact of our second-order Legendre truncation. When we extend the theoretical analysis to arbitrary multipole orders while maintaining the positivity constraint $q(\theta) \geq 0$, the optimal efficiency is achieved in the limit of delta-function-like localized heating at the rear pole of the particle (leading to $\beta \approx -3$, Supplementary Note 5.3)  analogous to the optimal source found in reference \cite{kreissl2016efficiency} for chemical swimmers, though identifying constraints \cite{kreissl2016efficiency}. This point-like heating naturally corresponds to a pusher mode, confirming that the constraint-induced selection of $\beta \approx -1$ is a fundamental feature of thermophoretic propulsion under physical realizability constraints, independent of the mathematical representation. Our second-order truncation captures this essential physics while remaining experimentally accessible—thermal diffusion and optical resolution limit the creation of perfectly localized heating in practice.

This finding suggests a broader principle: physical realizability constraints may universally shift optimal swimming modes away from hydrodynamic predictions. An intriguing open question is whether this $\beta \approx -1$ optimum is specific to spherical thermophoretic swimmers, or represents a more general feature. For non-spherical swimmers, three scenarios are possible: (1) \textit{Universal selection}—all phoretic swimmers under realizability constraints converge to $\beta \approx -1$ regardless of shape, with geometry affecting only absolute efficiency through resistance tensors; (2) \textit{Shape-dependent optima}—different geometries yield different optimal $\beta$ values as constraint-induced coupling competes with geometric effects; or (3) \textit{Constraint relaxation}—highly asymmetric shapes possess sufficient degrees of freedom to approach neutral swimming while satisfying $q \geq 0$. Our optical manipulation platform enables direct experimental testing of these scenarios by creating "effective shape asymmetries" through illumination pattern variations. Extending our analytical framework to arbitrary axisymmetric shapes requires solving the coupled optimization of power, temperature field, and shape under the combined constraints of heat equation, positivity, and fixed volume—an important direction for future theoretical work.

Beyond efficiency optimization, the demonstrated real-time reconfigurability enables functional versatility: microswimmers can reversibly switch between acting as propellers, pumps (generating force-dipole flows for shakers with $\beta = \pm\infty$), or stealth swimmers (with source-dipole-dominated flows for $\beta = 0$). This capability enables task optimization for specific environmental challenges, such as active morphing \cite{che2025arc}. Depending on the hydrodynamic swimming mode, the swimmer exhibits distinct hydrodynamic interactions with surfaces or other active components \cite{shen2018hydrodynamic, zottl2014hydrodynamics,rohde2025regulated}—pushers naturally clear paths by leaving empty space behind, while pullers create trailing structures. The ability to toggle between these modes on demand facilitates adaptive navigation in complex environments.

Our work opens the possibility of experimentally exploring the adaption and evolution of microswimmers in complex environments, their efficiencies and morphological developments which are highly dependent on external physical constraints. By demonstrating that physical constraints on the propulsion mechanis rather than geometry alone determine optimal swimming efficiency, our findings reveal a distinct pathway for evolutionary optimization in synthetic active matter: iterative refinement of programmable driving field patterns offers experimental access to constraint-guided adaptive dynamics without requiring morphological changes, complementing traditional shape-optimization approaches with a mechanism-optimization paradigm that may be more readily accessible in synthetic systems.

\section*{Methods}\label{methods}
\subsection*{Sample Preparation}
We use melamine formaldehyde (MF) particles (microParticles GmbH, radius a = $2.2\,\si{\Runit}$) coated with $10\,\mathrm{nm}$ diameter gold nanoparticles on the surface. For tracing the flow fields, we add gold nanoparticles (Sigma-Aldrich, diameter $250\,\mathrm{nm}$) to the sample solution. The sample consists of a thin water film confined between two microscope slides. The microswimmers are confined between the glass slides, thereby regulating the sample height. We seal the sample with polydimethylsiloxane to prevent evaporation. We observe particle motion with an inverted microscope (Olympus IX71) under dark field illumination using an oil-immersion dark field condenser (Olympus, NA = 1.2) and an oil-immersion objective (Olympus UPlanApo ×100/0.6). We record with a sCMOS camera (Hamamatsu) at an exposure time of $33 \,\si{\milli\second}$. We use particle image velocimetry (PIV) to extract velocities and obtain streamlines. We illuminate the microswimmers with a $\lambda = 532\,\mathrm{nm}$ laser. We generate different illumination patterns using a spatial light modulator (SLM).

\subsection*{Generation of laser light pattern}
We create distinct laser light profiles by shaping the laser light wavefront using a spatial light modulator (SLM). A $532\,$nm single-mode DPSS laser beam passing through a lambda half plate is expanded to illuminate the entire field of view of the SLM ($512\times512\,\mathrm{px}$). The SLM modulates the phase of the incoming linearly polarized light between $0$ and $2\pi$. A phase image of $N$ spatially distributed Gaussian spots is projected onto the SLM, which is positioned in the Fourier plane of the optical setup. We optimize both the number $N$ and the spatial arrangement of these Gaussian spots using a custom algorithm. During a Monte Carlo optimization, the goal is to generate an intensity profile that closely approximates the theoretical prediction shown in Supplementary Figure~S2 and S3. To quantify the match, we compute the mean squared error between the experimentally measured laser field and the theoretical intensity profile. The optimization terminates when this deviation falls below a predefined threshold of 0.001 or after a maximum of $40000$ iterations. More details on the algorithm are provided in Supplementary Note 2.\\
After scattering from the SLM surface, the light passes through a beam block positioned in the Fourier plane of the optical setup. This beam block removes the zeroth diffraction order originating from the pixelated structure of the SLM. An iris placed in the image plane crops the beam and suppresses unwanted illumination outside the target region. The pattern is then focused into the back focal plane of the oil-immersion objective (Olympus UPlanApo ×100/0.6), illuminating the sample plane.\\
To generate a moving thermo-phoretic microswimmer, we apply a steering method following previous studies \cite{fraenzl2021fully,wang2023spontaneous}. A custom LabVIEW program integrates all hardware components into a real-time feedback system. Camera-acquired images are directly fed into a Python program embedded in LabVIEW within approximately $20\,$ms. An intensity-based feature detection module analyzes the images and returns the $x$- and $y$-coordinates of detected particles within $5\,$ms. We compare the particle's coordinates to a predefined target position, which dictates the position and orientation of the light pattern projected onto the particle's surface to generate thermo-phoretic motion. In the next step, the phase mask of the light pattern is computed and sent to the SLM. This step also allows selection of the specific light pattern and thus the desired propulsion mode. In the consecutive camera frame, the particle's coordinates are reanalyzed and compared to the target, forming a feedback loop that updates the position and orientation of the projected light pattern.

\section*{Data Availability}
All data for display in the figure is available on request to the authors.

\section*{Code Availability}
The codes used in this study are available from the corresponding author upon reasonable request.

\section*{Acknowledgment}
L.~R. and F.~C acknowledge financial support from the German Research Foundation (Deutsche Forschungsgemeinschaft, DFG) through project no 432421051.

\section*{Author Information}
AUTHORS AND AFFILIATIONS\\
\textbf{Molecular Nanophotonics Group, Peter Debye Institute for Soft Matter Physics, Leipzig University}\\
Lisa Rohde, Gordei Anchutkin, Viktor Holubec, Frank Cichos\\
CONTRIBUTIONS\\
L.~R. and F.~C. designed the experiments. L.~R. carried out the experiments, simulations and model calculations. L.~R., V.~H., and F.~C., developed, discussed and verified the theoretical description. L.~R. and G.~A. developed the illumination pattern generation and G.A. developed the steering method. L.~R., V.~H., and F.~C. discussed the results and wrote the manuscript.

\section*{Competing interests}
The Authors declare no Competing Financial or Non-Financial Interests.

\section*{References}
\bibliographystyle{naturemag}
\bibliography{bib}
\end{document}